\documentclass[usenatbib]{mn2e}
\usepackage{graphicx,times}
\usepackage{url}
\usepackage{epstopdf} % my mac needs an explicit conversion

\begin{document}
\topmargin-1cm

%These look better, ADM thinks
\newcommand\approxgt{\lower.73ex\hbox{$\sim$}\llap{\raise.4ex\hbox{$>$}}$\,$}
\newcommand\approxlt{\lower.73ex\hbox{$\sim$}\llap{\raise.4ex\hbox{$<$}}$\,$}

%Additionally help simplify ADM's all-too-complex life
\def\Mpch {~h^{-1}~{\rm Mpc}}
\def\invMpch {~h~{\rm Mpc^{-1}}}

\title[Photometric clustering with full PDFs]{Incorporating Photometric
Redshift Probability Density Information into Real-Space Clustering
Measurements}
\author[A.~D.~Myers et al.]{Adam D. Myers$^1$\thanks{admyers@illinois.edu}, Martin White$^2$ and Nicholas M. Ball$^1$\\
${}^1$ Department of Astronomy, University of Illinois at Urbana- 
Champaign, Urbana, IL 61801\\
${}^2$ Department of Physics and Department of Astronomy, 601 Campbell Hall,
University of California Berkeley, CA 94720
}

\date{\today}
\pagerange{\pageref{firstpage}--\pageref{lastpage}}
\maketitle
\label{firstpage}

\begin{abstract}
The use of photometric redshifts in cosmology is increasing.  Often,
however these photo-$z$s are treated like spectroscopic observations,
in that the peak of the photometric redshift, rather than the full
probability density function (PDF), is used. This overlooks useful
information inherent in the full PDF. We introduce a new real-space
estimator for one of the most used cosmological statistics, the
2-point correlation function, that weights by the PDF of individual
photometric objects in a manner that is optimal when
Poisson statistics dominate. As our estimator does not bin
based on the PDF peak it substantially enhances the clustering signal
by usefully incorporating information from all photometric objects
that overlap the redshift bin of interest. As a real-world
application, we measure QSO clustering in the Sloan Digital Sky Survey
(SDSS). We find that our simplest binned estimator improves the clustering signal by a
factor equivalent to increasing the survey size by a factor of 2--3. We also introduce a new
implementation that fully weights between pairs of objects in constructing the cross-correlation
and find that this pair-weighted estimator improves clustering signal in a manner equivalent to 
increasing the survey size by a factor of 4--5. Our technique uses spectroscopic data to anchor
the distance scale and it will be particularly useful where
spectroscopic data (e.g, from BOSS) overlaps deeper photometry (e.g.,
from Pan-STARRS, DES or the LSST). We additionally provide simple,
informative expressions to determine when our estimator will be
competitive with the autocorrelation of spectroscopic objects.
Although we use QSOs as an example population, our estimator
can and should be applied to any clustering estimate that uses
photometric objects.

\end{abstract}


\section{Introduction} \label{sec:introduction}

With the advent of deep and wide multi-band photometric surveys there has been
a resurgence of interest in photometric redshifts as a means of estimating the
distance to a range of astrophysical objects.
Depending on the objects of interest and the information to hand, the derived
photometric redshifts will be of varying precision and accuracy, but all can be
described by a probability density function (PDF).
As our understanding of photometric redshifts improves our confidence in, and
ability to characterise, these PDFs, their use in cosmological statistical analyses is sure to increase.

In the sense that photo-$z$s represent color-redshift relations,
the use of an {\it ensemble\/} of PDFs for a {\it set\/} of objects is a
decades-old approach \citep[e.g.][and references therein]{Koo99}.
An example of this is the selection of cluster galaxies
\citep[e.g., via the Red Sequence;][]{Gla00}.
Cluster galaxy selection techniques have, in fact, recently been updated to
incorporate full PDFs \citep{vanB09} but approaches that use full PDFs remain
rare.
\citet{Sub96} introduced a method that used Gaussian PDFs to estimate
luminosity functions, a problem that has been studied for more arbitrary
PDF shapes by \citet{Che03} and \citet{Sheth07}.
Full PDFs are particularly underutilised in clustering work, where the use of
broad redshift bins is more prevalent.
By using broad redshift bins to measure photometric clustering one can
ameliorate uncertainties in the photo-$z$ ``peak", but typically at the
expense of constraining power.
%\citet{Bru00} used the ``peak" of the PDF to split angular clustering estimates
%across several broad redshift bins \citep[see also][]{Mye06,Mye07a,Mye07b}.

One of the most fundamental statistics of any population of objects, and one
which carries much physical information, is the 2-point correlation function
\citep[e.g.][]{Tot69}.
Provided the redshift distribution of the objects is well known, the underlying
3D clustering can be robustly inferred from the measured clustering in
projection \citep{Limber}, but the number of objects required increases
dramatically when the redshift distribution is broad.  For this reason,
estimates of the 2-point function can in principle gain tremendously from
improved utilization of the redshift information associated with photometric objects.

Often photo-$z$s are derived from the information in a subset of the objects
for which spectroscopy has been obtained.
In addition to calibrating the photo-$z$s, this subset of spectroscopic
objects can be used as distance anchors with which to set the real-space
transverse scale for distances to the photometric objects.
Measuring the cross-clustering of photometric objects around spectroscopic
objects has several advantages:
the properties of the spectroscopic objects, such as luminosity or spectral
type are precisely known;
the photometric objects are distributed more uniformly, meaning their
background clustering signature (the ``mask'') is simple to obtain and
issues like fiber collisions and more complex hidden selection dependencies that
might be introduced by the spectrographic setup are completely absent;
the cross-correlation probes the clustering only in a well-defined and
localised $z$-range, reducing the sensitivity to photometric outliers while
the number of pairs is dramatically increased by using the higher number
density of the photometric sample to improve statistics.
The use of spectroscopic-photometric cross-correlations to estimate clustering is not new \citep[e.g.][]{Lon79,Yee84,Yee87,Wol00,Hil91} however, using the information inherent in full PDFs to improve the
clustering signal in cross-correlation methods is in its infancy.

In this paper we develop a clustering measure which uses the full
photometric redshift PDF and which optimally weights
photometric-spectroscopic pairs in the limit that the error is
Poisson.  Our method circumvents the need to use the peak of the
photometric redshift PDF to select which objects lie in a redshift bin
of interest, or indeed to bin objects at all.  It allows every object
that can be assigned a photometric redshift to be usefully
cross-correlated against every spectrosopic object in the interval of
interest. We also provide simple, informative equations that indicate
when photometric redshifts are precise enough, for a given sample
size, to provide improved constraints over the spectroscopic
auto-correlation. We find that this condition is very hard to satisfy,
which explains why even relatively small spectroscopic surveys can
produce clustering measurements comparable to much larger photometric
samples. We additionally provide a quick method to calculate how much
our optimal weighting scheme for spectroscopic-photometric
cross-correlations can help satisfy this condition by using full PDF
information. The various equations we discuss should be very useful in
establishing a survey design to optimise clustering measurements.

To demonstrate our approach with real-world data we apply our new
method to measure the clustering of quasars (QSOs). The measurement of
QSO clustering sheds light on both QSO demographics and the physics
powering these systems.  The amplitude of clustering on large scales
is related to the masses of the dark matter halos which host the QSOs
(their environment), which together with the observed number density
allows QSO lifetimes or duty cycles \citep{ColKai89,HaiHui01,MarWei01}
to be constrained.  The small-scale clustering of QSOs can shed light
on their triggering mechanism, and on the nature of QSO progenitors.

With the advent of large, well-characterised samples, QSOs can now be
efficiently photometrically classified
\citep[e.g.][]{Ric04,Dab09,Ricopt09,RicIR09} but still have quite
imprecise photometric redshifts
\citep[e.g.][]{Bud01,Ric01,Wei04,Bal08}. This suggests that an
estimator that takes full advantage of the information in a
photometric redshift might be expected to dramatically improve
measurements of the clustering of QSOs. Most previous work on QSO
clustering used purely spectroscopic analysis
\citep{PorMagNor04,Cro05,PorNor06,Hen06,She07,Ang08,Mye08}, but all such
analyses are limited by the extremely low number density of objects
with spectra.  Higher number densities of objects can be achieved by
using photometric QSO selection \citep{Mye06,Mye07a,Mye07b} but
systematic errors must be carefully controlled because photometric
redshifts for QSOs are still frequently inaccurate. The use of
cross-correlations to measure QSO clustering has thus proven quite
popular \citep[e.g.][]{Cro04,AdeSte05a,AdeSte05b,Ser06,DEEP2,Str08,PWNP09,Mou09}. Our new technique
builds on such approaches, particularly that of \cite{PWNP09}, by
incorporating new information from photometric PDFs to improve the
clustering signal.

We note that, although we choose QSOs as our illustrative data set, our
methods and results are significantly more general and {\it our optimal
estimator will improve the signal for any real-space clustering measurement
that uses photometric redshifts}. Although the methods developed in this paper can be easily applied to any spectroscopic-photometric cross-correlation measurement, they will be of particular use in upcoming surveys where sparse spectroscopic data (e.g., from BOSS), is embedded in deeper photometric data, such as from PanSTARRS, DES and the LSST.

The outline of the paper is as follows.
\S\ref{sec:method} introduces our new optimal spectroscopic-photometric
cross-clustering estimator.
In \S\ref{sec:data} we introduce the QSO data we use as an example, and
in \S\ref{sec:qsoresults} we present the clustering results of this sample
and use it to demonstrate the improvement our new technique provides over
existing estimators that do not utilise the full PDF. 
We finish in \S\ref{sec:conclusions} with some conclusions and lessons
learned.
We assume a $\Lambda$CDM cosmological model with
$\Omega_{\rm m}=0.25$ and $\Omega_\Lambda=0.75$, consistent with the maximum likelihood estimates from the 5-year WMAP data \citep{Dun09}. All quoted magnitudes are corrected for Galactic extinction using the dust maps of \citet{Sch98}.

\section{Methodology} \label{sec:method}

\subsection{Real Space Clustering Measurements with Photometric Objects}
\label{sec:oldapproach}

Imagine we have a set of objects for which multi-band photometry has allowed
us to estimate photometric redshifts and a second (possibly disjoint)
set of objects for which spectroscopic redshifts are available.
For the spectroscopic objects we know (up to small uncertainties due to
peculiar velocities and uncertainties in the background cosmology) a physical
distance to each object, which can be used to anchor the physical scale.
Consider the cross-clustering between the set of objects with known
spectroscopic redshifts and the set of objects for which only photometric
redshifts are known.  To begin let us assume that the spectroscopic objects
all lie at a single redshift (and hence distance, $\chi_\star$) and relax
this assumption later.  We may estimate\footnote{More complex estimators,
such as that of \cite{LanSza93}, could also be used.  One would simply substitute each
estimator into Eq.~(\ref{eqn:cweightfinal}) or (\ref{eqn:enhanced}) evaluating
the $R_s(\chi_\star\theta)$ terms at different angular positions but at the
comoving distance of the spectroscopic data point.  We prefer the robustness
of Eq.~(\ref{eq:wtheta_DDDR}) to likely inaccuracies in the spectroscopic ``mask" over, e.g., the reduced variance
of the \cite{LanSza93} estimator.} the correlation function using the $DD/DR$
estimator \citep[e.g.][]{Sha83}
\begin{equation}
  w_{\theta}(R) = \frac{N_R}{N^{\rm phot}}\frac{D_sD_p(R)}{D_sR_p(R)} - 1 \,\,,
\label{eq:wtheta_DDDR}
\end{equation}
where we are measuring the cross-clustering of pairs of spectroscopic
and photometric objects, ``$D$'' denotes a data point ``$R$'' denotes
a point drawn from a random catalogue that mimics the data distribution
and the subscripts ``$p$'' and ``$s$'' denote ``photometric" and ``spectroscopic".
The factor $N_R/N^{\rm phot}$ scales the counts appropriately if the random
catalogue has a different size than the photometric catalogue.
We denote the random points $R_p$ both to specify that the random distribution
mimics the photometric data and to distinguish the term from $R=\chi_\star\theta$,
the transverse separation.
Note that Eq.~(\ref{eq:wtheta_DDDR}) only requires knowledge of the angular
selection function, or ``mask'', of the photometric data, not the
typically far more complex selection function of the spectroscopic data.
We have labeled this estimator $w_\theta(R)$ because it looks like a normal
angular correlation function in the photometric sample, except that angles
have been converted to distances using the distance to the spectroscopic
partner.

As detailed in \citet{PWNP09} we infer the projected, real-space,
cross-correlation function, $w_p(R)$, under the assumption that the
clustering is constant across the redshift slice and within the \citet{Limber}
approximation, using the relation
\begin{eqnarray}
  w_\theta(R) &=& \int d\chi\ f(\chi)
  \,\xi\left(R, \chi-\chi_\star\right) \\
  &\approx& f(\chi_\star) \int d\Delta\chi
  \ \xi\left(R, \chi-\chi_\star\right)\\
  &=& f(\chi_\star) w_{p}(R) \,\,,
\label{eq:wtheta}
\end{eqnarray}
where $f(\chi)$ is the normalised radial distribution function of the
photometric objects with $\int f(\chi)d\chi=1$ and all of the spectroscopic
objects lie at $\chi_\star$.
Note that this is a real space measurement and for broad enough $f(\chi)$
we can use the real-space correlation function in the integral, avoiding
the need to model redshift-space distortions.
Also note that we are making use of the fact that $f(\chi)$ is typically almost
constant across the entire line-of-sight range of integration employed in
defining $w_p$.  If this is not true then a more sophisticated analysis, which
factors in the changing selection function of ``random pairs'' with distance, is required.

For a distribution of spectroscopic redshifts one replaces $f(\chi_\star)$
in the above with the average, $\langle f(\chi_\star)\rangle$, across the
spectroscopic distribution.  For a small spectroscopic bin
($\chi_1\leq\chi<\chi_2$)
the redshift distribution will typically be flat.
In this case, $\langle f(\chi)\rangle$ tells us the fraction of objects in
the photometric data set that genuinely have redshifts in the spectroscopic
bin of interest ($f_z$) per comoving interval
($\langle f(\chi_\star)\rangle \approx f_z/(\chi_2-\chi_1)$.

We can use Eq.~(\ref{eq:wtheta}) to answer the question: how large does a
photometric sample need to be before a photometric-spectroscopic
cross-clustering measurement can compete with a spectroscopic auto-correlation?
Clearly, clustering estimates using photometric objects will improve as
photometric redshift precision (and accuracy) approaches the level of a
spectroscopic redshift (though in this limit our assumption of constant
$f(\chi)$ breaks down).
In the limit that the objects of interest are rare enough that their
clustering is dominated by Poisson shot-noise, then the angular bins
in $w_{\theta}(R)$ are independent and
\begin{equation}
  \frac{\delta w_\theta}{1+w_\theta} = N_{\rm pair}^{-1/2}
  \quad \Rightarrow \quad
  \frac{\delta w_p}{w_p} = \frac{f^{-1}+w_p}{w_p}\ N_{\rm pair}^{-1/2}
\label{eqn:dwp}
\end{equation}
where $N_{\rm pair}$ is the number of data pairs in the bin and $f$
is $\langle f(\chi_\star)\rangle$ for the photometric sample.  Note
that both $f^{-1}$ and $w_p$ have dimensions of length.
Eq.~(\ref{eqn:dwp}) neatly shows the main drawback of spectroscopic-photometric
cross-correlation measurements as compared to auto-correlation measurements
using only spectroscopic objects.  If the photometric redshift solutions are
significantly extended along the line-of-sight then $f_i$ is small (perhaps
as low as the reciprocal of the depth of the survey).  This suppresses the
measured clustering, $w_\theta$, which for a given sample is proportional to
$f$.  A very large number of pairs are thus necessary to measure $w_\theta$
with any precision.

How large is the typical suppression?
When measuring the spectroscopic auto-correlation the clustering is
integrated along the line-of-sight to eliminate the effects of
redshift-space distortions.  The limits of integration tend to vary
from author to author but typically the line-of-sight interval is
$\mathcal{O}(100\,h^{-1}\,{\rm Mpc})$.  In the language of
Eq.~(\ref{eqn:dwp}) such an auto-correlation estimate can approach a
limit of $f\approx 0.01\,h\,{\rm Mpc}^{-1}$.  If the photometric
sample is extended over, say, $1\,h^{-1}$Gpc, then
$f=\mathcal{O}(10^{-3}\,h\,{\rm Mpc}^{-1})$, and the number of
photometric objects needs to be larger by a factor of $\sim100$ in
order to measure the clustering as well as if precise redshifts were
known.
If the extent is $500\,h^{-1}$Mpc one needs $\sim 25$ times more
objects, and for $300\,h^{-1}$Mpc one needs $\sim 10$ times as many.
Of course, if obtaining spectroscopy or improved PDFs for the photometric
sample is unrealistic then one has no other choice but to use the existing
information.

\subsection{An Optimal Estimator for Real-Space Clustering using
Photometric Redshifts}

\begin{figure}
\begin{center}
\resizebox{3.2in}{!}{\includegraphics{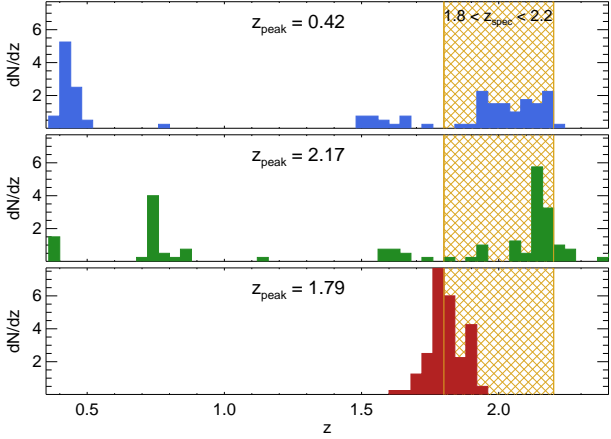}}
\end{center}
\caption{In analyses that use the PDF peak, only the PDF in the centre panel ($z_{\rm peak}=2.17$) would be considered to overlap the spectroscopic bin of interest ($1.8 < z_{\rm spec} < 2.2$ in this plot). In reality each PDF has a 50\% overlap with the spectroscopic bin. We illustrate some typical problems with using PDF peaks; PDFs that overlap the spectroscopic bin but have a preferred peak solution far from the bin (a ``catastrophic" redshift; upper panel), PDFs with a peak solution in the bin but that are smeared out across a large range of redshifts (centre panel), and well-defined PDFs that lie just outside the bin of interest (lower panel). The PDFs are for real photometric QSOs calculated using the method of \citet{Bal08}.}
\label{fig:problempdfs}
\end{figure}

We have noted two major drawbacks to measuring the real-space clustering of
photometrically classified objects around spectroscopic objects.
First, it is not clear how to establish which photometric objects should
be cross-correlated with a given set of spectroscopic objects.
The typical approach would be to use objects with a peak photometric redshift
solution in the redshift bin of interest.
This, however, discards much of the information codified in the photometric
redshift PDF and ignores the fact that an object with a peak photometric
redshift in the range of interest may actually have less chance of being in
that redshift range than an object with a peak photometric redshift beyond
that range, particularly as the peak of the PDF may itself be poorly defined.
We illustrate this in Figure~\ref{fig:problempdfs}.
The second drawback is the possible extension of the ensemble of the
photometric redshifts along the line-of-sight, which causes $f$ to be
small in Eq.~(\ref{eqn:dwp}).

We now introduce a new method designed to circumvent these issues.
Consider breaking the photometric sample into very thin slices in photometric
redshift, $z_p$, and labelling the slices from $i=1,\cdots,k$. Each photometric sample, 
$i$, provides an estimate of $w_p(R)$ via
$w_\theta(R)/f_i$.  Writing this estimate as $w_i(R)$, with an error
proportional to $f_i^{-1}N_{\rm pair}^{-1/2}$ in the limit of weak
clustering, we can inverse variance weight the different measurements
to obtain
\begin{equation}
  w_p(R) = \sum_i N_i^{\rm phot} f_i^2 w_i(R) \Bigm\slash
           \sum_i N_i^{\rm phot} f_i^2
\label{eqn:wpvarweight}
\end{equation}
where $N^{\rm phot}_i$ is the number of photometric objects in sample $i$.
This circumvents the issue of which photometric objects to
cross-correlate against a set of spectroscopic objects in a chosen bin
of redshift.  Clearly photometric samples which peak at very different
redshifts from the spectroscopic sample are significantly
down-weighted in the sum.  Note that our method also down-weights both
objects with unusual colours that might have multi-peaked PDFs and
objects with poorly constrained photometry, such as near survey
limits, where the PDF might be very broad.

Since the binning is so far arbitrary we can consider the limit where each
slice in Eq.~(\ref{eqn:wpvarweight}) represents a single photometric object,
i.e.~$N_i^{\rm phot}=1$ for each $i$.
In this case photometric objects that have some overlap with the spectroscopic
bin of interest are included in the sum and photometric objects with zero
overlap have zero weight.
Treating the photometric objects individually, rather than in an ensemble,
removes the need for any arbitrary binning and effectively reduces the
extension of the ensemble PDF along the line-of-sight and should thus
significantly improve the clustering signal-to-noise.

\begin{figure}
\begin{center}
\resizebox{3.2in}{!}{\includegraphics{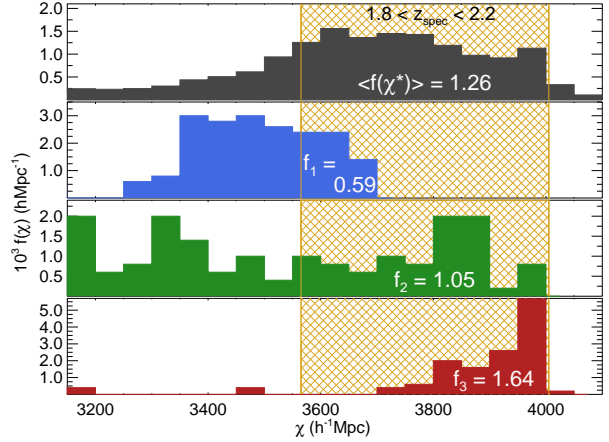}}
\end{center}
\caption{The calculation of $\langle f(\chi_\star)\rangle$ and $f_i$, the
``comoving overlaps'', in units of $10^{-3}\invMpch$. The upper panel
demonstrates the old method (\S\ref{sec:oldapproach}), in which the
photometric redshift PDFs are combined into $\langle
f(\chi_\star)\rangle$ an ensemble, normalised, comoving distribution
averaged over the spectroscopic bin of interest ($1.8 < z_{\rm spec} <
2.2$ in this plot). The lower panels demonstrate our new bin-weighted estimator
(Eq.~\ref{eqn:cweightfinal}) in which each PDF is transformed into a
normalised comoving distribution and averaged across the bin of
interest $f_1, f_2, f_3...f_k$. The lower panels displays the case for
$N_i^{\rm phot}=1$ in Eq.~(\ref{eqn:wpvarweight}) but any number $N^{\rm phot}$ of PDFs can be
combined into an ensemble.}
\label{fig:chibin}
\end{figure}

Because the weights in Eq.~(\ref{eqn:wpvarweight}) are
$\sigma_i^{-2} = N_i^{\rm phot} f_i^2$
a rough determination of how much this new estimator will improve the
signal-to-noise of a $w_p$ estimate over existing methods, which only
consider objects that have a peak photometric redshift in the bin of
interest is
\begin{equation}
  \sum_i N_i^{\rm phot}f_i^2 \Bigm\slash n\langle f(\chi_\star)\rangle^2
\label{eqn:comp}
\end{equation}
where the $i$ subscripts represent our new optimal estimator for a
slice containing $N^{\rm phot}$ photometric objects and the $n$ represents the
number of photometric objects with a PDF peak in the
spectroscopic bin of interest. The $f_i$ are the comoving fractional
photometric redshift overlaps for objects in slice $i$ and $\langle
f(\chi_\star)\rangle$ is the same for the ensemble of photometric
objects with a peak photometric redshift in the spectroscopic bin of
interest. This is illustrated in Figure~\ref{fig:chibin}, in which the upper panel
plots the ensemble of the ($n=110410$) PDFs with $1.8 < z_{\rm peak} < 2.2$. 
This ensemble has an $\langle f(\chi_\star)\rangle=1.26\times10^{-3}\invMpch$ overlap with the true range $1.8 < z < 2.2$. 
The lower panels plot three individual (i.e. $N_1^{\rm phot}=N_2^{\rm phot}=N_3^{\rm phot}=1$) PDFs 
and their overlaps with $1.8 < z < 2.2$.

\subsection{The Optimal Estimator in Practice}

In \S\ref{sec:qsoresults}, we illustrate the degree to which our optimal estimator
can improve clustering estimates for a ``typical'' analysis, using a sample
of spectroscopic and photometric QSOs.
QSOs may be particularly well suited to our estimator as they are rare
enough that their clustering is dominated by Poisson noise (e.g., see Figure~\ref{fig:bootstrap}) out to reasonably
large scales and $f(\chi)$ is quite broad.
We note, though, that our optimal estimator should improve the signal-to-noise
for any photometric clustering analysis.
The exact methodology we use in practice is as follows.
Eq~(\ref{eqn:wpvarweight}) can be rewritten as
\begin{equation}
  w_p(R) = \sum_i c_i w_i^{\theta}(R)
\label{eqn:cweight}
\end{equation}
\noindent where
\begin{equation}
  c_i = N_i^{\rm phot} f_i \Bigm\slash
           \sum_i N_i^{\rm phot} f_i^2
\label{eqn:fweight}
\end{equation}
and we have used $w_p=w_\theta/f_i$.
Now, consider substituting Eq.~(\ref{eq:wtheta_DDDR}), the typical $DD/DR$
estimator for $w(\theta)$, into Eq.~(\ref{eqn:cweight})
\begin{equation}
  w_p(R) = \sum_i c_i \left[\frac{N_R}{N_i^{\rm phot}}\,\frac{D_sD_p(R)}{D_sR_p(R)} -
  1\right]
\label{eqn:wpcweight}
\end{equation}
where the the transverse separation, $R$, is evaluated using the angle
between a spectroscopic-photometric pair and the distance to the
spectroscopic object.  Finally we obtain a simple equation for
calculating the real-space clustering of a sample of photometric
objects with full PDFs around a sample of spectroscopic objects
\begin{equation}
  w_p(R) = N_R\sum_i \frac{c_i}{N_i^{\rm phot}}\,\frac{D_sD_p(R)}{D_sR_p(R)} -
  \sum_ic_i \quad .
\label{eqn:cweightfinal}
\end{equation}
The $1/N_i^{\rm phot}$ factor reflects the fact that care must be taken to weight
the random catalogue correctly, i.e., on a slice-by-slice basis. Note that $\sum c_i\sim
f^{-1}(\chi_\star)$ approximates the reciprocal of $\langle
f(\chi_\star)\rangle$ from the unweighted estimator. We prefer
Eq.~(\ref{eqn:cweightfinal}) to other versions of this expression as
it facilitates simple tracking of the data-data counts to construct
error estimates from subsampling of the counts.

Finally, we note that one can express the weights in Eq.~(\ref{eqn:fweight}) based on overlaps between each individual spectroscopic and photometric object (i.e. weighting fully by pairs rather than by how much a photometric object overlaps a {\em bin} of many spectroscopic objects) without loss of generality. The equations of interest would then reduce to 

\begin{equation}
  c_{i,j} = N_i^{\rm phot}N_j^{\rm spec}f_{i,j} \Bigm\slash
           \sum_{i,j} N_i^{\rm phot}N_j^{\rm spec}f_{i,j}^2
\label{eqn:enhancedvarweight}
\end{equation}

\noindent where $N_j^{\rm spec}$ is the number of spectroscopic objects in slice $j$. We will choose $N_j^{\rm spec} = 1$ (as well as $N_i^{\rm phot} = 1$) throughout. Similarly

\begin{equation}
  w_p(R) = N_RN_s\sum_{i,j} \frac{c_{i,j}}{N_i^{\rm phot}N_j^{\rm spec}}\,\frac{D_sD_p(R,\Delta\chi)}{D_sR_p(R,\Delta\chi)} -
  \sum_{i,j}c_{i,j} \quad 
\label{eqn:enhanced}
\end{equation}

\begin{figure}
\begin{center}
\resizebox{3.2in}{!}{\includegraphics{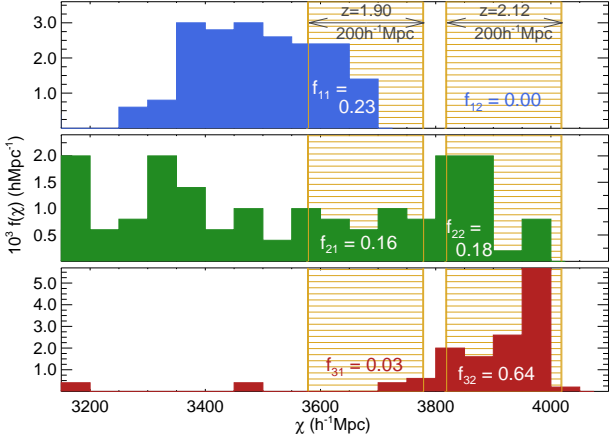}}
\end{center}
\caption{The calculation of $f_{ij}$, the
``comoving overlaps'' for the pair-weighted approach of Eq.~(\ref{eqn:enhanced}). A comoving window ($\Delta\chi\pm100\Mpch$ in the case of this plot) is adopted around each spectroscopic QSO, which are indexed $j$. There will be many spectroscopic QSOs in a given redshift bin of interest but here we plot only two at $z=1.90$ and $z=2.19$ for illustrative purposes. Each photometric PDF, indexed $i$, is then averaged across each of the comoving windows to produce pairs of weights $f_{ij}$. We display the case for
$N_i^{\rm phot}=N_j^{\rm spec}=1$ in Eq.~(\ref{eqn:enhancedvarweight}) but any number $N^{\rm phot}$ of PDFs and $N^{\rm spec}$ of spectroscopic slices can be combined into ensembles.}
\label{fig:pairweight}
\end{figure}

\noindent where $N_s$ is the total number of spectroscopic objects analyzed in the spectroscopic bin of interest and $\Delta\chi$ is the size of the comoving window integrated over around each spectroscopic object. The additional normalization of $N_s$ arises by analogy with Eq.~(\ref{eqn:cweightfinal}) and the addition of new spectroscopic slices. The extent of the comoving window is entirely flexible, requiring some trial-and-error to determine the optimal choice, although $\Delta\chi\sim\mathcal{O}(50$--$100\,h^{-1}\,{\rm Mpc})$, as used when integrating out the spectroscopic autocorrelation to eliminate
the effects of redshift-space distortions, is an obvious choice. This slightly enhanced approach should provide additional
signal-to-noise gains over Eq.~(\ref{eqn:cweightfinal}) provided the photometric PDFs are sufficiently sampled 
to accurately estimate their overlap with small comoving distance intervals. We illustrate this final, full pair-weighted approach in Figure~\ref{fig:pairweight}.

\section{Data} \label{sec:data}

Although our main result is the new methodology outlined
in \S\ref{sec:method}, in \S\ref{sec:qsoresults} we will
illustrate our new method with real-world samples to demonstrate
the improvements that it can return.  We will
make use of quasars selected from the SDSS, as described here.

\subsection{Photometric Quasars} \label{sec:KDE}

The photometric quasar sample that we analyze is constructed using the
Kernel Density Estimation (KDE) technique of \citet{Ric04},
a technique to classify quasars in photometric surveys which draws
on several innovations inherent to the SDSS (e.g., \citealt{York00}) --
extensive and carefully monitored $ugriz$ imaging
(e.g., \citealt{Gun98,Hog01}) calibrated to a standard photometric system
(e.g., \citealt{Fuk96,Smi02}) with a precision of a few-hudredths of a
magnitude \citep{Ive04}. These innovations allow quasars to be more easily separated
{}from the stellar locus.
We use the DR6 KDE sample, which is detailed in full in \citet{Ricopt09}.

The DR6 KDE sample is drawn from a test sample of all point sources in the
SDSS DR6 imaging data \citep{DR6} with $i <21.3$, where $i$ refers to the
{\em asinh} magnitude \citep{Lup99} in the ``uber-calibrated'' system of \citet{uber}.
The DR6 primary imaging data covers an area of $8417\,{\rm deg}^2$ but further
cuts \citep{Mye06,Ricopt09} remove approximately $150\,{\rm deg}^2$ or $1.7$\% of
the area.

In this paper we concern ourselves only
with DR6 KDE objects that have a very high probability of being QSOs. 
As such, we apply a \textit{uvxts=1} cut within the sample. This cut selects QSOs at particularly high efficiency by limiting the DR6 KDE sample to QSOs that would have been selected by traditional UV-excess techniques. As noted in Table~4 of \citet{Ricopt09}, and discussed in \citet{Mye06}, only
$\sim$5\% of  the \textit{uvxts=1} QSOs should,
in reality, be stars\footnote{\citet{Ricopt09} advocate a  \textbf{good~$\ge$~0} cut to improve efficiency. We ignore this, as for 
\textit{uvxts=1} it only discards a further 2.4\% of the
data.}. The UV-excess nature of the \textit{uvxts=1} cut limits the spectroscopic redshift range to $0.8~\approxlt~z~\approxlt~2.4$.

\begin{figure}
\begin{center}
\resizebox{3.2in}{!}{\includegraphics{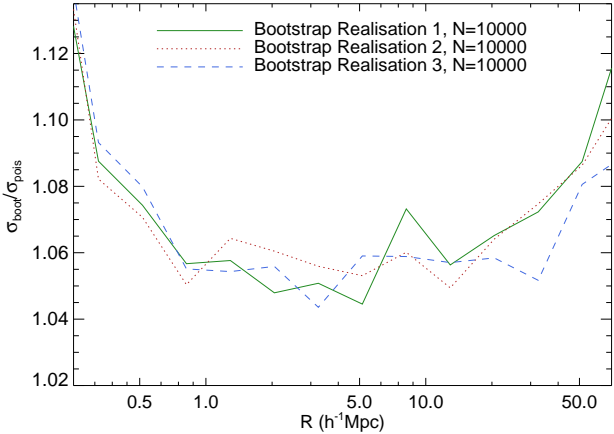}}
\end{center}
\caption{The ratio of the bootstrap error to the Poisson error for the
old, ensemble method of \S\ref{sec:oldapproach}. We plot three
separate realizations to demonstrate that the error is stable to
$\sim1$\% for 10,000 bootstraps. The bootstrap error tracks the
Poisson error to around 6\%. On scales $\approxlt~0.5\Mpch$, where
there are few QSO pairs, 10,000 bootstraps is insufficient to recreate
the shot noise. On scales $\approxgt~20\Mpch$, where QSO pairs are not
independent, Poisson errors underestimate the true error. This plot
demonstrates that bootstrapping (at N=10,000) and Poisson errors agree
well in the range $0.5 < R < 20\Mpch$.}
\label{fig:bootstrap}
\end{figure}

\setlength{\tabcolsep}{5.58pt}
\begin{table}
\centering
\begin{tabular*}{0.4703\textwidth}{|c|cccccccc|}
\hline
$z$ & 0.8 & 1.0 & 1.2 & 1.4 & 1.6 & 1.8 & 2.0 & 2.2 \\ \hline
Imp. & 1.87 & 1.61 & 1.22 & 1.63 & 1.53 & 1.40 & 1.77 & 1.90 \\ \hline
(Imp.)$^2$ &  3.50 & 2.60 & 1.48 & 2.65 & 2.35 & 1.96 & 3.15 & 3.63 \\ \hline
\end{tabular*}
\caption{\small{``Imp.'' is the expected improvement due to our new
method (Eq.~\ref{eqn:cweightfinal}) over the old ensemble approach
(\S\ref{sec:oldapproach}) as characterised by Eq.~(\ref{eqn:comp}). As
this value approximates the improvement in Poisson noise, its square
approximates the equivalent increase in survey size.}}
\label{table:expec}
\end{table}

\subsubsection{Redshift Distribution of Photometric Quasars}
\label{sec:dndz}

While estimating the redshift of a QSO with a large number of narrow
filters can be precise (e.g., \citealt{Hat00,Wol01,Wol03})
results using broadband filters are more mixed (e.g.,
\citealt{Ric01,Bud01}). Although photometric redshifts are often
expressed as a single value, they are, in reality,
probabilistic, with a full probability density function (or PDF)
representing the possible redshifts the object of interest could
occupy given the filter information. Our main goal in this paper is
to incorporate full PDF information into clustering
analyses. If we denote by $P_s^j(z)$ the probability density function
for QSO $j$, and assume $\int P_s^j(z)dz=1$ across all possible redshifts, 
then the value that will interest us is the fraction
of the ensemble PDF that will genuinely lie in any redshift
interval $z_1<z<z_2$
\begin{equation}
  f_z = \frac{1}{N^{\rm phot}}\sum_{j=1,N^{\rm phot}} \int_{z_1}^{z_2}dz\ P_s^j(z) \quad .
\label{eqn:f}
\end{equation}
This fraction can be deduced for arbitrary redshift intervals and could correspond to a single photometric QSO ($N^{\rm phot} =1$) having, say, a 60.3\%
chance of lying in the redshift range of interest, or equivalently a
sample of 100 PDFs in an ensemble from which we might derive that 60.3
of the 100 QSOs in the ensemble can be expected to actually lie in the interval of
interest.

We obtain our PDFs using the Nearest Neighbour approach outlined in
\citet{Bal08}.  We perturb a QSO's colours relative to a spectroscopic training set drawn from the DR5 QSO sample \citep{DR5QSO}, determine the nearest neighbour over 100 perturbations, and
build a function that describes the probability that the photometric quasar
matches near spectroscopic neighbours.\footnote{Our PDFs for the DR6KDE catalog will be made available at \url{http://lcdm.astro.uiuc.edu/nbckde_dr6_pdfs}} Examples of these PDFs are shown in Figures~\ref{fig:problempdfs} and \ref{fig:chibin}.

\subsection{Spectroscopic Quasars}

We cross correlate the above QSOs with a sample of spectroscopic QSOs
drawn from the DR6 QSO sample (Schneider et al. 2009 in prep, see
\citealt{DR5QSO}). Our spectroscopic QSO sample populates the sky in a
complex manner but for our method, only the distribution of the
photometric sample, which is far simpler, needs to be modeled.

We impose the criterion that our spectroscopic QSOs must also appear
in the photometric sample discussed in \S\ref{sec:KDE}. We make no
additional cuts on flags or redshift quality, as the vast majority of
quasar redshifts are reliable if the object is, indeed, a QSO, and the
cuts made by \citet{Ricopt09} help ensure both the quality of the
photometry of the QSO, and the likelihood that it is a QSO.

\section{Example Implementation of the New Optimal Estimator}
\label{sec:qsoresults}

In this section, we apply the method developed in \S\ref{sec:method}
to the spectroscopic and photometric QSO samples discussed in
\S\ref{sec:data} to illustrate both our new methodology and its statistical
gains over current methods. As our goal is a simple
demonstration of our new methodology, we apply no cuts to the samples
beyond those discussed in \S\ref{sec:data}. This ensures that any
improved signal is due to the method itself, rather than any additional
magnitude, colour or redshift cuts that we might impose. As outlined in
\S\ref{sec:data}, the only significant cut we employ is the
\textit{uvxts=1} cut within the photometric sample. This cut, which is
purely to ensure that almost all of our photometric objects are genuinely QSOs, limits our spectroscopic redshift range to
$0.8~\approxlt~z~\approxlt~2.4$.

\subsection{Expected Improvement in Signal}

Eq.~(\ref{eqn:comp}) allows us to estimate how treating each
photometric QSO's PDF individually (i.e. Eq.~\ref{eqn:cweightfinal})
will improve the clustering signal over treating the photometric QSOs
as an ensemble (as discussed in \S\ref{sec:oldapproach}). In
Figure~\ref{fig:chibin} we demonstrate the calculation of $\langle
f(\chi_\star)\rangle$ for two different approaches; the ensemble
approach of \S\ref{sec:oldapproach} and our new bin-weighted approach
(Eq.~\ref{eqn:cweightfinal}), which treats each $f_i$ individually. In
Table~\ref{table:expec} we show the expected improvement implied by
Eq.~(\ref{eqn:comp}) for a range of spectroscopic redshift bins. This
improvement arises from using all of the information inherent in every
PDF for every individual photometric object and is about a
factor of $\sim1.6\times$. Based on Poisson
statistics, simply using our new approach should be roughly
equivalent to having a $\sim2$--$3\times$ larger survey.

\subsection{Actual Improvement in Signal}

Poisson errors are typically used to calibrate the noise in a clustering estimator (e.g.,
\citealt{LanSza93})
\begin{equation}
 \Delta w_{\theta}(R) = \frac{1+ w_{\theta}(R)}{\sqrt{D_sD_p(R)}}
\label{eq:poiserr}
\end{equation}
Poisson errors accurately reflect the clustering noise on small scales
(where many pairs remain independent) and remain very accurate for the
photometric sample being used out to at least $20\Mpch$ (e.g.,
consider deprojecting Figure~1 of \citealt{Mye06}). Poisson errors are
more complex to calculate for our new methodology because we
incorporate pairs of points with unequal weights, some that may be
completely outside the spectroscopic bin of interest, but they can in
principle be computed.  However we estimate the errors by simply bootstrapping
\citep[e.g.,][]{Efron} on the individual {\em spectroscopic\/} QSOs, as was
done in \citet{PWNP09}.  This approach is additionally useful as it
demonstrates how one might estimate errors for our new approach based
on other resampling approaches, such as jackknifes or field-to-field
variations. Resampling approaches are generally more accurate than
Poisson errors on large scales and facilitate the construction of a
full covariance matrix. Our preferred expressions for our new estimators
(Eq.~\ref{eqn:cweightfinal} and \ref{eqn:enhanced}) make it straightforward to track how
each {\em spectroscopic\/} QSO affects the pair counts and quickly
construct resampled error estimates.

\begin{figure}
\begin{center}
\resizebox{3.2in}{!}{\includegraphics{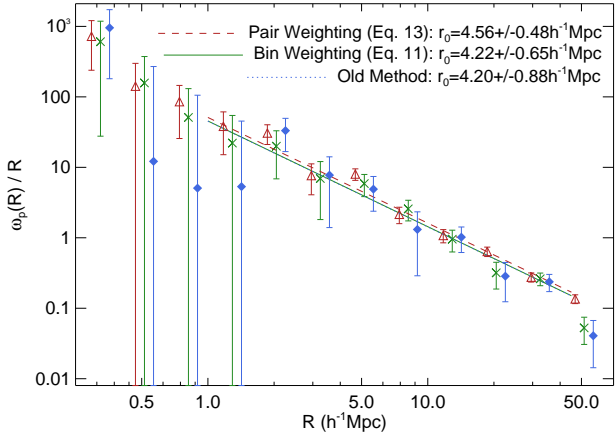}}
\end{center}
\caption{$w_p(R)$ as measured by the old, ensemble estimator (diamonds; Eq.~\ref{eq:wtheta}) and
our new bin-weighted estimator (crosses; Eq.~\ref{eqn:cweightfinal}) and full pair-weighted estimator (triangles; Eq.~\ref{eqn:enhanced}). The pair-weighted estimator for this plot used a comoving window of $\pm50\Mpch$. All plotted data are for QSOs with spectroscopic redshifts in the bin $1.8 < z_{\rm spec} < 2.2$. We fit a $\gamma=1.5$ power law over $1.6 <R<40\Mpch$ to each estimate using the full covariance matrix estimated from 10,000 bootstraps. The points have been offset slightly for
display purposes. The best fit value of the comoving scale length
$r_0$ (see Eq.~\ref{eq:powerlaw}) is displayed for each data set,
together with the ($2\sigma$) error on the fit.}
\label{fig:firstresult}
\end{figure}

In Figure~\ref{fig:bootstrap} we plot the relationship between the
Poisson and bootstrap errors derived for the ensemble estimator (i.e.,
derived using only QSOs with peak PDF solutions in the spectroscopic
bin of interest, as discussed in \S\ref{sec:oldapproach}) using a
spectroscopic bin of $1.8 \leq z_s < 2.2$. Across scales of $0.2 < R <
50\Mpch$ the bootstrap errors converge to within $\sim0.8$\% for
10,000 bootstraps, and the amplitude of the bootstrap errors closely
tracks (within $\sim5$--10\%) that of the Poisson errors. This
demonstrates that bootstrapping on the spectroscopic QSOs is close to
equivalent to using Poisson errors on the scales of interest. On
scales $\approxlt~0.5\Mpch$, where there are few QSO pairs, more
bootstrap samples are likely needed to recreate the precision of the
Poisson errors. On scales $\approxgt~20\Mpch$ the Poisson errors
likely begin to underestimate the noise as covariance increases.

Having demonstrated the validity of bootstrapping to obtain estimates
of the noise we plot the results for the old
ensemble approach, our new bin-weighted estimator (Eq.~\ref{eqn:cweightfinal}) and our full pair-weighted estimator (Eq.~\ref{eqn:enhanced}) in Figure~\ref{fig:firstresult}. To summarise our
results we fit power laws to our data. A power-law 3D
correlation function of the form $\xi(r)=(r/r_0)^{-\gamma}$ produces a
power-law projected correlation function
\begin{equation}
  \frac{w_{p}(R)}{R} = \frac{\sqrt{\pi}
  \,\Gamma[(\gamma-1)/2]}{\Gamma[\gamma/2]} 
  \left(\frac{r_0}{R}\right)^{\gamma}
  \quad .
\label{eq:powerlaw}
\end{equation}
We fit this form to the measured correlations over the range $1.6
<R<40\Mpch$, using the full bootstrap covariance and holding the index
fixed at $\gamma=1.5$. In order to improve the numerical stability of
this procedure, we scale $w_p(R)$ by $R^{1/2}$, thereby removing the
artificially high condition number that arises due to the large
dynamic range of $w_p$. The power-law fit for the old, ensemble,
approach gives $r_0 = 4.20 \pm 0.88$, our new bin-weighted estimator (Eq.~\ref{eqn:cweightfinal}) gives $r_0 =
4.22 \pm 0.65$ ($2\sigma$) and our full pair-weighted method (Eq.~\ref{eqn:enhanced}) gives $r_0 =
4.56 \pm 0.48$ ($2\sigma$), which agree well with numerous recent
estimates of the amplitude of $w_p$ for QSO clustering near $z\sim2$
(e.g., \citealt{PorMagNor04,Cro05,PorNor06,Ang08}). We list $2\sigma$
errors to reflect the fact that our errors are likely underestimated
on large scales but the relative improvements for our new estimators are
identical whether we quote $1\sigma$ or $2\sigma$ errors.

\setlength{\tabcolsep}{4.6pt}

\begin{table}
\centering
\begin{tabular*}{0.4703\textwidth}{|c|cccccccc|}
\hline
$R$ & \multicolumn{8}{c|}{$z$} \\
$(\Mpch)$ & 0.8 & 1.0 & 1.2 & 1.4 & 1.6 & 1.8 & 2.0 & 2.2 \\ \hline 
% $0.5\Mpch$ &  1.44 &  1.45 &  1.14 &  1.27 &  1.26 &  1.13 &  1.32 &  1.30 \\ hline
 0.8 &  1.41 &  1.25 &  1.08 &  1.29 &  1.30 &  1.25 &  1.39 &  1.36 \\ 
 1.3 &  1.43 &  1.28 &  1.11 &  1.34 &  1.27 &  1.21 &  1.35 &  1.44 \\  
 2.0 &  1.41 &  1.28 &  1.11 &  1.29 &  1.27 &  1.18 &  1.39 &  1.44 \\  
 3.2 &  1.41 &  1.27 &  1.12 &  1.28 &  1.26 &  1.22 &  1.38 &  1.43 \\ 
 5.1 &  1.42 &  1.29 &  1.13 &  1.30 &  1.26 &  1.21 &  1.36 &  1.43 \\ 
  8.2 &  1.38 &  1.30 &  1.11 &  1.33 &  1.27 &  1.19 &  1.34 &  1.40 \\ 
12.9 &  1.41 &  1.30 &  1.10 &  1.30 &  1.27 &  1.20 &  1.35 &  1.43 \\   
20.5 &  1.38 &  1.28 &  1.11 &  1.28 &  1.26 &  1.20 &  1.33 &  1.36 \\ \hline
10.5 & 1.36  & 1.28  & 1.11  & 1.25  & 1.25  & 1.20  & 1.33  & 1.44  \\ \hline
%$32.5\Mpch$ &  1.34 &  1.27 &  1.13 &  1.28 &  1.26 &  1.18 &  1.36 &  1.40 \\ hline  
\end{tabular*}
\caption{\small{Improvement of our new bin-weighted estimator (Eq.~\ref{eqn:cweightfinal}) over the old methodology of \S\ref{sec:oldapproach}. Each column represents a bin width of 0.4 in (spectroscopic) redshift centred on $z$. The scales in the first column are logarithmic at five-per-decade. Table values are the ratio between jackknife errors for the new to the old estimator ($\sigma_{\rm new}/\sigma_{\rm old}$). The final row is the total improvement over $1 < R < 20\Mpch$. Squaring the table values approximates the equivalent increase in survey size obtained by using our estimator.}}
\label{table:actual}
\end{table}

It is clear from the fits and errorbars in
Figure~\ref{fig:firstresult} that our new bin-weighted estimator (Eq.~\ref{eqn:cweightfinal}), which utilises all
of the redshift information in the PDF not just the peak of the PDF,
considerably improves the signal-to-noise in estimates of $w_p(R)$. In
Table~\ref{table:actual} we list the improvement in signal-to-noise as
a function of redshift and scale for our sample. Our new bin-weighted estimator,
across scales that are typically used to represent the quasi-linear
regime of clustering ($1 < R < 20\Mpch$) improves the signal-to-noise
of clustering estimates by 30\%. Adopting our
most basic approach of incorporating full PDFs into a clustering measurement is
thus equivalent to increasing the size of the photometric sample
discussed in \S\ref{sec:KDE} by 60\%.  Photometric redshift
determinations for QSOs in broadband $ugriz$ are particularly poor
outside of the range $1 < z < 2$.  Outside of this range, the
improvement yielded by our bin-weighted estimator is slightly larger, equivalent to
increasing the survey size by 80\%.

We note that our improvements in Table~\ref{table:actual} are slightly
smaller than the expected improvements listed in Table~\ref{table:expec}.
This could reflect a breakdown in our assumption of Poisson errors or
innaccuracy in our PDFs.  In fact, one novel approach of our methodology
would be to tune the PDFs until the figures in Table~\ref{table:actual}
peaked, thus constructing PDFs without using any colour information
(see also \citealt{Sch06}).

In Table~\ref{table:pairweight} we list the improvement in signal-to-noise as
a function of scale using our full pair-weighted estimator (Eq.~\ref{eqn:enhanced}) for a spectroscopic redshift bin of $1.8 < z < 2.2$.
We adopt a representative range of comoving windows (see the discussion of $\Delta\chi\sim\mathcal{O}(50$--$100\,h^{-1}\,{\rm Mpc})$ near Eq.~\ref{eqn:enhanced}). The improvement in signal-to-noise is about a factor of 2 for scales that are typically used to represent the quasi-linear regime of clustering ($1 < R < 20\Mpch$). Across some scales the improvement in signal approaches a factor of $2.2\times$ for a comoving window of $\Delta\chi=\pm50\Mpch$. Impressively, this means that our full pair-weighted estimator can potentially improve clustering by a factor equivalent to increasing the size of a survey by a factor of 4--5.

The improvements in Tables~\ref{table:actual} and \ref{table:pairweight} demonstrate that
the PDFs we use must carry additional information that can be used to improve
clustering signal, which was the main goal of this paper. In future, as our
knowledge of PDF construction is refined, the improvements facilitated by
our method can only also improve.

\setlength{\tabcolsep}{9.9pt}

\begin{table}
\centering
\begin{tabular*}{0.439\textwidth}{|c|c|ccc|}
\hline
 $R$ & Eq.~(\ref{eqn:cweightfinal}) &  \multicolumn{3}{c|}{Eq.~(\ref{eqn:enhanced}); $\Delta\chi$ in $\Mpch$} \\
$(\Mpch)$ &  & $\pm200$ & $\pm100$ & $\pm50$ \\ \hline
 0.8 & 1.39 &   1.41 &  1.76 &  2.03 \\          
 1.3 & 1.35 &   1.39 &  1.80 &  2.10 \\          
 2.0 & 1.39 &   1.43 &  1.79 &  2.10 \\          
 3.2 & 1.38 &   1.44 &  1.81 &  2.16 \\          
 5.1 & 1.36 &   1.42 &  1.76 &  2.05 \\          
 8.2 & 1.34 &   1.42 &  1.79 &  2.16 \\          
12.9 & 1.35 &   1.39 &  1.77 &  2.11 \\          
20.5 & 1.33 &   1.34 &  1.70 &  1.99 \\  \hline
10.5 & 1.33 &   1.34 &  1.68 &  2.04 \\ \hline
\end{tabular*}
\caption{\small{Improvement of our full pair-weighted estimator (Eq.~\ref{eqn:enhanced}) over the old methodology of \S\ref{sec:oldapproach} and our binned estimator (Eq.~\ref{eqn:cweightfinal}). Each calculation is over a spectroscopic bin of $1.8 < z < 2.2$. Table values are the ratio between jackknife errors for the new estimators as compared to the old estimator ($\sigma_{\rm new}/\sigma_{\rm old}$). For the full pair-weighted estimator (Eq.~\ref{eqn:enhanced}) the columns are the adopted comoving window around each spectroscopic QSO. The equivalent window for Eq.~(\ref{eqn:cweightfinal}) would be $\sim220\Mpch$, corresponding to the full bin $1.8 < z < 2.2$. The final row is the total improvement over $1 < R < 20\Mpch$. Squaring the table values approximates the equivalent increase in survey size obtained by using our estimators.}}
\label{table:pairweight}
\end{table}

\section{Conclusions} \label{sec:conclusions}

We have introduced new correlation function estimators to improve measurements of how photometric objects cluster around spectroscopic objects. Spectroscopic-photometric cross-correlations have known benefits, due to the spectroscopic objects having narrowly-defined distance information and the photometric objects having significantly higher number densities. Our approach uses the full photometric probability density information, or PDFs, to optimise such cross-correlation estimates in the Poisson limit. We note that It is possible that a strict Poisson weighting for pairs can be improved upon, particularly on moderate scales. 

We have additionally provided simple equations that can be used to calculate when our new estimators will improve on measurements from the spectroscopic autocorrelation. The parameters of interest are the overlap of the photometric data with the spectroscopic bin in comoving space, which depends on the PDF precision, and the relative number of photometric and spectroscopic objects. Because the number of photometric objects scales as the square of the the comoving overlap it can be difficult for spectroscopic-photometric cross-correlations to improve on spectroscopic autocorrelation estimates.

Our improved estimator has several benefits over existing cross-correlation methods. Most obviously, because our estimator does not solely rely on the ``peak'' of a photometric object's PDF to determine which photometric objects should be cross-correlated against the spectroscopic objects of interest, the information from more photometric objects is used in clustering estimates. We show that, in the case of photometric QSOs, simply using the bin-weighted form of our estimator (Eq.~\ref{eqn:cweightfinal}) can thus improve signal-to-noise in the Poisson limit in a manner equivalent to obtaining almost $2\times$ as much survey data. Eq.~(\ref{eqn:comp}) suggests that the full gains on all scales may be closer to equivalent to obtaining $3\times$ as much survey data. Indeed, our full pair-weighted estimator Eq.~(\ref{eqn:enhanced}) demonstrates that gains equivalent to increasing survey size by as much as a factor of 4--5 can be realised. Although we have specifically used the example of QSOs, we stress that our estimator can and should be used to improve the signal for any real-space clustering measurement using photometric redshifts. 

The current incarnation of our method has several shortcomings. If the PDFs peak sharply relative to the spectroscopic redshift distribution then $f(\chi)$ cannot be validly extracted, and the full integration across Eq.~(2) must be applied. Our assumptions similarly break down if the spectroscopic survey selection function varies rapidly across the redshift bin of interest. In these cases the full 2D correlation function must be integrated in the line-of-sight direction. These inadequacies cannot be countered by narrowing the spectroscopic bin indefinitely, as redshift-space distortions ultimately limit the scale where redshifts map to line-of-sight distances. As such, our assumptions are most robust for the pair-weighted methodology of Eq.~(\ref{eqn:enhanced}). In this pair-weighted approach, a strict spectroscopic window of, say, $\pm50\Mpch$ can be enforced, and our assumptions would then be valid until the PDFs are more precise than $\pm50\Mpch$ or the spectroscopic distribution varies rapidly over $\pm50\Mpch$.

A particular benefit of our estimator is that it can, very simply, incorporate {\em every} photometric object into an analysis, negating the need to bin the photometric objects. PDFs of varying precision from a range of photometric data can thus be simply combined in a single measurement, provided the mask of photometric object {\em detections} is well-controlled. One could thus envisage taking, say, multi-wavelength photometry from patchy space telescope data or a range of small dedicated surveys (to improve PDFs where possible) embedded in uniform optical photometry such as the SDSS (to establish detections of the photometric objects of interest), and straightforwardly cross-correlating this complex photometric data with a completely different spectroscopic data set. Further, there is no reason to limit the probabilistic information to a photometric redshift. Many techniques, such as star-galaxy separation or the star-QSO separation technique we have used in this paper \citep{Ricopt09}, provide classification probabilities as well as photometric redshifts. Such classification probabilities can naturally be incorporated into our method by, e.g., weighting a PDF heavily to $z=0$ if an object has a high probability of being a star.

Because of the flexibility of our estimator, it should be useful anywhere on the sky where spectroscopic data is embedded in deep, potentially complex and multi-wavelength, photometric data. This should make our estimator particularly useful for regions of the sky where extensive spectroscopy, such as from BOSS, the various 2dF surveys and the SDSS, is embedded in deep, well-calibrated photometry, with measurable PDFs such as from Pan-STARRS, DES and the LSST. Over the next decade, we expect that obvious applications of our estimator will include improved measurements of the clustering of photometric LBGs, LRGs and QSOs around spectroscopic QSOs and measuring the clustering of photometric galaxies and QSOs around absorption features in QSO spectra.

\section*{Acknowledgements}

ADM was supported in this work by NASA ADP grant NNX08AJ28G and by the
University of Illinois. MW is supported by NASA and the DOE. We thank
Gordon Richards, Alex Gray, Robert Nichol and Robert Brunner for their
substantial and important work in helping produce the KDE photometric
QSO catalogue and Nikhil Padmanabhan and Britt Lundgren for helpful conversations.

Funding for the SDSS and SDSS-II has been provided by the Alfred P. Sloan
Foundation, the Participating Institutions, the National Science Foundation,
the U.S. Department of Energy, the National Aeronautics and Space
Administration, the Japanese Monbukagakusho, the Max Planck Society, and
the Higher Education Funding Council for England.
The SDSS Web Site is {\tt http://www.sdss.org}.

The SDSS is managed by the Astrophysical Research Consortium for the
Participating Institutions.  The Participating Institutions are the
American Museum of Natural History, Astrophysical Institute Potsdam,
University of Basel, University of Cambridge, Case Western Reserve University,
University of Chicago, Drexel University, Fermilab, the Institute for
Advanced Study, the Japan Participation Group, Johns Hopkins University,
the Joint Institute for Nuclear Astrophysics, the Kavli Institute for
Particle Astrophysics and Cosmology, the Korean Scientist Group,
the Chinese Academy of Sciences (LAMOST), Los Alamos National Laboratory,
the Max-Planck-Institute for Astronomy (MPIA), the Max-Planck-Institute
for Astrophysics (MPA), New Mexico State University, Ohio State University,
University of Pittsburgh, University of Portsmouth, Princeton University,
the United States Naval Observatory, and the University of Washington.

\label{lastpage}
\end{document}